\documentclass[prd,aps,12pt,nofootinbib,superscriptaddress,showpacs]{revtex4-1}

\usepackage{amssymb,amsthm,amsbsy}

\newcommand{\beq}{\begin{equation}}
\newcommand{\eeq}{\end{equation}}

\newcommand{\ber}{\begin{eqnarray}}
\newcommand{\eer}{\end{eqnarray}}

\def\cD{{\cal D}}

\def\ri{{\rm i}}
\def\rg{{\rm g}}
\def\rP{{\rm P}}

\begin{document}

\title{Post-inflationary preheating with weak coupling}

\author{Igor Rudenok}
\affiliation{Department of Physics, Taras Shevchenko National University, Kiev, Ukraine} %

\author{Yuri Shtanov}
\affiliation{Bogolyubov Institute for Theoretical Physics, Kiev 03680, Ukraine} %
\affiliation{Department of Physics, Taras Shevchenko National University, Kiev, Ukraine} %

\author{Stanislav Vilchinskii}
\affiliation{Department of Physics, Taras Shevchenko National University, Kiev, Ukraine} %
\affiliation{D\'{e}partement de Physique Th\'{e}orique and Center for Astroparticle
Physics, Universit\'{e} de Gen\`{e}ve, 24 Quai Ernest-Ansermet, CH-1211, Gen\`{e}ve 4, Switzerland}%

\begin{abstract}
Particle production in the background of an external classical oscillating field is a key
process describing the stage of preheating after inflation.  For sufficiently strong
couplings between the inflaton and matter fields, this process is known to proceed
non-perturbatively.  Parametric resonance plays crucial role for bosonic fields in this
case, and the evolution of the occupation numbers for fermions is non-perturbative as
well. In the Minkowski space, parametric resonance for bosons and non-perturbative
effects for fermions would still persist even in the case of weak coupling. In
particular, the energy density of created bosons would grow exponentially with time.
However, the situation is quite different in the expanding universe.  We give a simple
demonstration how the conditions of the expanding universe, specifically, redshift of the
field modes, lead to the usual perturbative expressions for particle production by an
oscillating inflaton in the case of weak couplings. The results that we obtain are
relevant and fully applicable to the Starobinsky model of inflation.
\end{abstract}

\pacs{98.80.Cq}

\maketitle

\section{Introduction}

Reheating is one of the most important epoch of the universe evolution, connecting the
inflation stage with the subsequent hot Big-Bang phase. Almost all matter constituting
the universe at the subsequent radiation-dominated stage was created during the reheating
stage. In most models of inflation based on the scalar field (inflaton), the universe is
usually preheated by particle creation in the background of an oscillating inflaton.  The
particles created are subsequently thermalized, and the universe becomes hot. Initially,
this process was treated perturbatively, e.g., by using the Born approximation for the
decay rates of the inflaton (see \cite{Linde:1990}). later, it was realized
\cite{Traschen:1990sw, KLS, Shtanov:1994ce} that creation of bosons may also proceed
non-perturbatively via the effect of parametric resonance, and that the creation of
fermions is non-perturbative as well \cite{GK}.  This is especially true for sufficiently
strong couplings between the inflaton and other fields, in which case the resonance is
broad in the frequency space \cite{KLS}.

%One  should notice  that the question: ``Is the Born formula (perturbation theory)  a
%limiting case of the parametric resonance regime?'' has an answer ``No, the standard
%effect of scalar field decay described by the perturbation theory and preheating due to
%parametric resonance are two different effects and the perturbative decay of the scalar
%field may coexist with the parametric resonance".

When the coupling between the inflaton and other fields is sufficiently weak, then, in
some typical cases, it is considered legitimate to return to the usual Born perturbation
theory in calculating the particle production rates {\em in the expanding universe\/}
(see, e.g., \cite{Shtanov:1994ce}). This may look somewhat puzzling if one takes into
account that the effect of parametric resonance for bosons, as well as the
non-perturbative evolution of the occupation numbers of fermions, would certainly occur
in the background of a classical field oscillating {\em in the Minkowski space-time\/},
however small is the coupling.  In this case, therefore, the Born formula would not be
applicable. Why, then, does it work in the space-time of an expanding universe?
We clarify this issue in the present paper. We give a simple demonstration as to how the
conditions of the expanding universe, notably, the redshift of frequencies of the field
modes, result in the usual perturbative expressions for particle production by an
oscillating inflaton in the case of weak couplings.

In this paper, we consider a model of inflation based on a power-law potential for the
inflaton field, and the inflaton field  throughout this paper will be regarded as
classical. In Sec.~\ref{sec:prelim} we describe  the particle production via parametric
resonance in the Minkowski space for the case in which  a scalar field $\phi$ of mass $M$
is weakly interacting (interaction constant $\zeta$ is sufficiently small) with a light
scalar field $\varphi$ of mass $m_\varphi \ll M$.  We then describe several important
modifications in the case of expanding universe, where $\phi$ plays the role of the
inflaton, and particle creation becomes an essential part of the preheating process. In
some cases (violation the adiabaticity condition), parametric resonance does not develop
in the expanding universe, and one returns to the Born approximation for the total width
$\Gamma_\varphi$ of decay of the field $\phi$ into a pair of $\varphi$ particles, which
wold not be valid in the Minkowski space.  In Secs.~\ref{sec:bose} and \ref{sec:fermi},
we derive the Born approximation in the case of expanding universe, for bosons and
fermions, respectively.  In Sec.~\ref{sec:starobinsky}, we show that the results obtained
are relevant and fully applicable to the Starobinsky model of inflation
\cite{Starobinsky:1980te}.  In Sec.~\ref{sec:inflaton}, we study the issue of the
inflaton self-production in this model.  Our general conclusions are formulated in
Sec.~\ref{sec:conclude}.

\section{Preliminaries}
\label{sec:prelim}

Consider a scalar field $\phi$ of mass $M$ interacting with a light scalar field
$\varphi$ of mass $m_\varphi \ll M$ with the interaction Lagrangian density
\beq \label{scoup}
{\cal L}_{\rm int} = - \zeta \phi \varphi^2 \, ,
\eeq
where $\zeta$ is a constant with dimension mass.  As our initial conditions, the
homogeneous field $\phi (t)$ is classically oscillating in the neighborhood of its
minimum at $\phi = 0$ with amplitude $\phi_0$, while the field $\varphi$ is in the vacuum
state. In the Minkowski space, this situation would lead to particle production via
parametric resonance.  Specifically, for sufficiently small values of $\zeta$, namely,
for
\beq \label{smalls}
\zeta \phi_0 \ll M^2 \, ,
\eeq
the resonance will be most efficient in the first narrow resonance band centered at the
frequency
\beq
\omega_{\rm res} = \frac{M}{2}
\eeq
(see \cite{Traschen:1990sw, KLS, Shtanov:1994ce}).  Within the resonance band, the mean
particle occupation numbers grow with time according to the law
\beq \label{resn}
N_k = \frac{1}{1 - \Delta^2 / \zeta^2 \phi_0^2} \sinh^2 \lambda\, t \, ,
\eeq
where
\beq \label{lambda}
\lambda = \frac{1}{M} \sqrt{\zeta^2 \phi_0^2 - \Delta^2} \, , \qquad \Delta = \omega_k^2
- \omega_{\rm res}^2 \, ,
\eeq
and $\omega_k = \sqrt{m_\varphi^2 + k^2} \approx k$ is the frequency of the mode of the
field $\varphi$. The width of the resonance band of frequencies $\omega_k$ is determined
by the condition that the expression under the square root in (\ref{lambda}) is
nonnegative.

The total particle number, as well as the energy density of the $\varphi$-particles, in
Minkowski space grows asymptotically exponentially with time, in contrast to the
expectations based on the na\"{\i}ve perturbation theory, where it grows with time only
linearly.

There are several important modifications in the case of expanding universe, where $\phi$
plays the role of the inflaton, and particle creation is an essential part of the
preheating process \cite{Traschen:1990sw, KLS, Shtanov:1994ce}.  Firstly, the amplitude
of the oscillating inflaton gradually decreases with time as $\phi_0 \propto a^{-3/2}$,
where $a$ is the scale factor. Secondly, the frequency of the mode of the scalar field
$\varphi$ is redshifted:
\beq \label{omk}
\omega_k = \sqrt{m_\varphi^2 + \frac{k^2}{a^2}} \approx \frac{k}{a} \, ,
\eeq
where $k$ now is the comoving wave number, and we took into account that the mode is
close to the resonance, hence, its wave number is relativistic.

One should note the difference of the picture in the Minkowski space and in the expanding
universe.  In the Minkowski space, the frequency picture is static. The modes within the
resonance band will stay there all the time, and the resonant enhancement of the created
particle number will occur in such modes. Thus, the resulting occupation numbers will
grow according to (\ref{resn}). In the expanding universe, however, the frequency picture
is dynamically changing: the frequencies of the modes of the field to be created are
evolving due to the redshift.  Each particular mode spends a finite time in the resonance
zone, so that only a limited gain of the occupation number is possible. It is not clear
{\em a priori\/} whether the resonant growth will occur or not in the expanding universe.

The theory of parametric resonance is applicable if the evolution of the relevant
quantities occurs adiabatically. Specifically, if
\beq \label{phiad}
\left| \frac{\dot \phi_0}{\phi_0} \right| = \frac32 H \ll M \, ,
\eeq
where $H \equiv \dot a / a$ is the Hubble parameter, and if
\beq \label{muad}
\left| \frac{\dot \lambda}{\lambda} \right| \ll \lambda \, ,
\eeq
then one can replace law (\ref{resn}) by an approximate expression \cite{Shtanov:1994ce}
\beq
N_k \simeq \sinh^2 \int \lambda\, d t \, ,
\eeq
as long as the mode with the comoving wave number $k$ remains within the resonance band.

If the adiabaticity condition (\ref{muad}) does not hold, and the parametric resonance,
therefore, does not develop, then one usually employs the Born approximation for the
total width $\Gamma_\varphi$ of decay of a $\phi$ particle into a pair of $\varphi$
particles:
\beq \label{Gammab}
\Gamma_\varphi = \frac{\zeta^2}{8 \pi M} \, .
\eeq
However, if this na\"{\i}ve formula does not work in the Minkowski space (as argued
above), one may wonder why it works in the case of expanding universe, with continuously
redshifted particle momenta etc.

Similar issues can be raised about the production of fermionic particles.  Although there
is no parametric resonance in this case, still the picture of creation of particle pairs
by an oscillating {\em classical\/} field is quite different from that based on the usual
perturbation theory \cite{GK}.  Nevertheless, in the case of expanding universe, one
often uses the Born formula for the total width of decay of $\phi$ into a pair $\overline
\psi, \psi$\,:
\beq \label{Gammaf}
\Gamma_\psi = \frac{\Upsilon^2 M}{8 \pi}
\eeq
where $\Upsilon$ is the Yukawa coupling of the scalar field $\phi$ to the fermionic field
$\psi$.

The widths (\ref{Gammab}) and (\ref{Gammaf}) in the Born approximation in the background
of an oscillating classical field in the Minkowski space are calculated, e.g., in
\cite{Shtanov:1994ce}.

The purpose of this paper is to clarify the formulated issues and to show how equations
(\ref{Gammab}) and (\ref{Gammaf}) arise in the case of expanding universe.  We then show
that these equations are fully applicable to the Starobinsky model of inflation
\cite{Starobinsky:1980te}.

\section{Bosons}
\label{sec:bose}

A scalar field $\varphi$ with mass $m_\varphi$ interacting with the inflaton $\phi$ via
coupling (\ref{scoup}) obeys the equation of motion
\beq \label{eq-varphi}
\Box \varphi + \left( m_\varphi^2 + 2 \zeta \phi \right) \varphi = 0 \, .
\eeq

Turning to the mode components $\varphi_k$ one gets the equation
\beq
\ddot{\varphi}_k+3H\dot{\varphi}_k+\left(\frac{k^2}{a^2} + m_\varphi^2 + 2 \zeta \phi
\right) \varphi = 0
\eeq
%Proceeding to the conformal time $\eta$ so that the metric has the form
%\beq
%ds^2 = d t^2 - a^2 (t) d x^2 = a^2 (\eta) \left( d \eta^2 - d x^2 \right) \, ,
%\eeq
In order to eliminate the friction term, it is convenient to rescale the field as $\chi_k
= a^{3/2} \varphi_k$. For the mode $\chi_k$ with the comoving wave number $k$, at the
preheating stage, we have the equation (see, e.g., \cite{Shtanov:1994ce})
\beq \label{basic}
\ddot \chi_k + \Omega^2_k \chi_k = 0 \, ,
\eeq
where
\begin{equation} \label{Omegak}
\Omega_k^2 (t) = \omega_k^2 (t) + 2 \zeta \phi(t) - \frac94 H^2 - \frac32 \dot H \, ,
\end{equation}
and $\omega_k$ is given by (\ref{omk}).  At the preheating stage, the inflaton field
evolves as
\begin{equation} \label{oscill}
\phi (t) = \phi_0 (t) \cos M t \, ,
\end{equation}
where the amplitude $\phi_0 (t) \propto a^{-3/2} (t)$ slowly decreases with time due to
the universe expansion as a consequence of the adiabaticity condition (\ref{phiad}).

The equation for the Hubble parameter in an inflaton-dominated universe is
\beq \label{rhophi}
H^2 = \frac{1}{3 M_\rP^2} \rho_\phi \, , \qquad \rho_\phi = \frac12 M^2 \phi_0^2 \, ,
\eeq
where
\beq
M_\rP = \left( 8 \pi G \right)^{-1/2} = 2.4 \times 10^{18}\,\mbox{GeV}
\eeq
is the reduced Planck mass.  Under conditions (\ref{phiad}), (\ref{muad}), the boson
particle production proceeds via the effect of parametric resonance, as described in the
preceding section.  In this case, the last two terms in (\ref{Omegak}) can be neglected.

In this paper, we are interested in the case where condition (\ref{muad}) is {\em
violated\/}, so that parametric resonance does not have time to develop and plays no
role. Using equation (\ref{lambda}), we see that, in the model under consideration,
violation of condition (\ref{muad}) in the center of the resonance band is equivalent to
\beq \label{smallss}
| \zeta |  \lesssim \sqrt{\frac38} \frac{M^2}{M_\rP} \, .
\eeq
Since the inflaton-field oscillations mainly occur in the regime $\phi_0 \ll M_\rP$, we
see that condition (\ref{smalls}) is, in fact, a consequence of this inequality.

The effect of non-stationarity of the external field $\phi$ and of the metric is that the
quantity $\Omega_k$ in the equation of motion (\ref{basic}) is a function of time. In the
case $\Omega_k={\rm const}$, the solution for $\varphi_k$ would maintain its
positive-frequency character, i.e., $\varphi_k \sim e^{\ri \Omega_k t}$ for all $t$. The
time-dependence of $\Omega_k$ results in the mixing of frequencies, hence, in particle
production of the field $\varphi$.  Under condition (\ref{smallss}), parametric resonance
does not play any role, and the particle occupation numbers are small.  Hence, in
calculating them, one is justified to use perturbation theory.

The mixing of frequencies is considered in a standard way by looking for solutions of the
field equation in the form \cite{Grib:1994}
\ber
\varphi_k (t) &=& \frac{1}{\sqrt{\Omega_k}} \left[ \alpha_k (t) e^{\ri \int^t_{t_0}
\Omega_k(t') d t'} + \beta_k (t) e^{- \ri \int^t_{t_0} \Omega_k (t') d t'} \right] \, , \label{phiq} \\
\dot \varphi_k (t) &=& \ri \sqrt{\Omega_k} \left[ \alpha_k (t) e^{\ri \int^t_{t_0}
\Omega_k (t') d t'} - \beta_k (t) e^{-\ri \int^t_{t_0} \Omega_k (t') d t'} \right] \, ,
\label{dotphiq}
\eer
where $\alpha_k (t)$ and $\beta_k (t)$ are the Bogolyubov coefficients satisfying the
relation
\beq
\left| \alpha_k \right|^2 - \left| \beta_k \right|^2 = 1 \, .
\eeq
In terms of these coefficients, the average occupation numbers in the corresponding modes
are given by $N_k = |\beta_k|^2$. Thus, to find the number of created particles, one
needs to find the coefficient $\beta_k$. Substituting expressions (\ref{phiq}),
(\ref{dotphiq}) into (\ref{basic}), one obtains the following system of equations for
$\alpha_k$ and $\beta_k$\,:
\ber
\dot{\alpha}_k &=& \frac{\dot \Omega_k}{2 \Omega_k} e^{- 2 \ri \int^t_{t_0} \Omega_k (t')
d t'} \beta_k \, , \\
\dot \beta_k &=& \frac{\dot \Omega_k}{2 \Omega_k} e^{2 \ri \int^t_{t_0} \Omega_k (t') d
t'} \alpha_k  \, .
\eer
The initial conditions for these equations are $\alpha_k = 1$, $\beta_k = 0$.  Then,
treating this system perturbatively, in the first order, the equation for coefficient
$\beta_k$ is
\begin{equation}
\dot \beta_k (t) = \frac{\dot \Omega_k}{2 \Omega_k} e^{2 \ri \int^t_{t_0} \Omega_k (t') d t'} \, .
\end{equation}

Using equation (\ref{Omegak}) and employing the adiabaticity approximation (\ref{phiad}),
one transforms this equation into
\begin{equation}
\dot \beta_k = \left( \frac{\dot \omega_k}{2 \omega_k} + \frac{\zeta M \phi_0}{2 \omega_k^2} \sin
M t \right) e^{2 \ri \int^t_{t_0} \omega_k (t') d t'} \, .
\end{equation}
Expressing the function $\sin M t$ as a sum of exponents, we get
\begin{equation}
\dot \beta_k = \frac{\dot \omega_k}{2 \omega_k} e^{2 \ri \int^t_{t_0} \omega_k (t') d t'}
+ \frac{\zeta M \phi_0}{4 i \omega_k^2}   e^{2 \ri \int^t_{t_0} \omega_k (t') d t'+i M
t} + \frac{\zeta M \phi_0}{4 i \omega_k^2}   e^{2 \ri \int^t_{t_0} \omega_k (t') d t'-i
M t} \, .
\end{equation}

The first two terms oscillate with time with high frequency and thus do not give
appreciable contribution to the coefficient $\beta_k$. The third term might pass through
the resonance leading to the gain of $\beta_k$. Leaving only theis resonant term, one
obtains an approximate solution for the Bogolyubov coefficient $\beta_k$ in the form
\begin{equation} \label{beta}
\beta_k = \frac{\ri \zeta M}{4} \int_{t_0}^t \frac{\phi_0 (t')}{\omega_k^2 (t')} e^{2 \ri \int^{t'}_{t_0}
\omega_k (t^{\prime\prime}) d t^{\prime\prime} - \ri M t'} d t' \, .
\end{equation}
As most particles are created in a narrow resonance region of frequencies, we can extend
the limits of integration in (\ref{beta}) to infinity and use the stationary-phase
approximation to estimate the value of integral. This gives
\beq \label{ocnumb}
\beta_k = \frac{\zeta M \phi_0 (t_k)}{4 \omega_k^2 (t_k)} \sqrt{\frac{\pi}{\left|
\dot \omega_k (t_k)\right|}} = \frac{\zeta \phi_0 (t_k)}{M} \sqrt{\frac{\pi}{\left| \dot
\omega_k (t_k)\right|}} = \frac{\zeta \phi_0 (t_k)}{M^{3/2}} \sqrt{ \frac{2 \pi}{H
(t_k)}} \, ,
\eeq
where the moment of time $t_k$ is defined by the stationary-phase relation $\omega_k
(t_k) = M /2$, which is just the moment of passing through the center of the resonance
band for the $k$-mode.

One can picture the process of particle creation in the following way.  A mode with
sufficiently high wave number $k$ undergoes redshift till it reaches the resonance
region.  After passing through the narrow resonance band, it becomes filled with
particles with average occupation numbers (\ref{ocnumb}), which, in our approximation,
remains subsequently constant.  In this picture, the instantaneous particle spectrum is
given by the following approximation:
\begin{equation}
N_k = \left\{
\begin{array}{rl} 0\, , &\ \ \displaystyle k > \frac{M a (t)}{2} \quad
\mbox{(the $k$-mode has not yet passed through resonance)} \, , \medskip \\
|\beta_k|^2\, , &\ \ \displaystyle k_{\rm min} < k < \frac{M a (t)}{2} \quad
\mbox{(the $k$-mode has already passed through resonance)} \, , \medskip \\
0\, , &\ \ \displaystyle k < k_{\rm min} \quad \mbox{(the $k$-mode will never
pass through resonance)} \, .
\end{array}
\right.
\end{equation}
All modes with momentum less than $k_{\rm min} = M a (t_0) / 2$, where $t_0$ is the
moment of the beginning of particles creation, will never pass through the resonance
region due to the redshift.

In this picture, the energy density $\rho_{\varphi}(t)$  of the created particles at any
moment of time is given by
\begin{equation}
\rho_\varphi (t) = \frac{1}{a^4 (t)} \int \frac{d^3 k}{(2 \pi)^3}\, \theta \left( k -
k_{\rm min} \right) \theta \left(M a(t) - 2 k \right) k |\beta_k|^2 \, ,
\end{equation}
where $\theta (x)$ is the Heaviside step function.

The effective rate of particle production $\Gamma_\varphi$ is determined by comparing the
time derivative of this energy density with the appropriate equation for the evolution of
the energy density $\rho_\varphi$ of continuously created relativistic particles
\beq
\dot \rho_\varphi = - 4 H \rho_\varphi + \Gamma_{\varphi} \rho_{\phi} \, .
\eeq
We have
\begin{equation}
\Gamma_{\varphi} \rho_{\phi} = \left. \frac{k^3|\beta_k|^2 M \dot a (t)}{4 \pi^2 a^4
(t)} \right|_{k = M a(t)/ 2} \, ,
\end{equation}
whence, using (\ref{rhophi}) and (\ref{ocnumb}), we get the standard expression
(\ref{Gammab}) for the quantity $\Gamma_\varphi$.

\section{Fermions}
\label{sec:fermi}

In a curved space-time, one uses the covariant generalization of the Dirac equation:
\begin{equation} \label{Dirac0}
\left[ \ri \gamma^{\mu}(x) \cD_{\mu} - m \right] \psi(x) = 0 \, .
\end{equation}
Here, $\gamma^{\mu}(x)=h^{\mu}_{(a)} (x)\gamma^a$, where $\gamma^a$ are the usual Dirac
matrices, and the tetrad vectors $h^{\mu}_{(a)} (x)$ are normalized as
\begin{equation}
h^\mu_{(a)}h^{}_{(b)\mu}=\eta_{ab} \, ,
\end{equation}
with $\eta_{ab}$ being the metric of the flat space.

%Let us introduce the dual tetrad vectors $h^{(a)}_\mu$ normalized by the condition
%$h^{(a)}_\mu h^{\mu}_{(b)}=\delta^a{}_b$, so that
%\begin{equation}
%h^{(a)}_\mu h_{(a)\nu}=g_{\mu\nu} \, .
%\end{equation}

The covariant derivative  $\cD_{\mu}$ of the Dirac field in (\ref{Dirac0}) is defined as
\begin{equation}
\cD_\mu\psi=\left[\partial_\mu+\frac14C_{abc}h^{(c)}_{\mu}\gamma^b\gamma^a\right]\psi \, .
\end{equation}
Here, $C_{abc}$ are the Ricci coefficients, which are related to the tetrad vectors by
\begin{equation}
C_{abc}=\left(\partial_\nu h^{\mu}_{(a)}\right)h_{(b)\mu}h^\nu_{(c)}\, .
\end{equation}

We consider the standard case where the spinor field $\psi$ interacts with the inflaton
field $\phi$ through the Yukawa coupling
\beq \label{Yukawa}
{\cal L}_{\rm int} = \Upsilon \phi \overline \psi \psi \, .
\eeq
This results in the appearance of effective time-dependent fermion mass in equation
(\ref{Dirac0}):
\beq \label{varmass}
m (t) = m_{\psi} - \Upsilon \phi(t) \, .
\eeq

It is convenient to proceed to the conformal time coordinate $\eta=\int a^{-1}(t)dt$, in
terms of which the metric becomes conformally flat:
\beq
ds^2=a^2(\eta) \left( d\eta^2 - dl^2 \right)\, .
\eeq
In cosmological setting, the tetrad vectors in the spherical coordinates can be chosen in
the form
\begin{equation}
\begin{array}{ll}
h_{(0)0} = -h_{(1)1}=a(\eta)\, , &h_{(2)2}=-a(\eta)f(\chi)\, , \\
h_{(3)3} = - a (\eta) f(\chi) \sin{\theta}\, , \quad &h_{(a)i}=0\, , \quad a\neq i \, .
\end{array}
\end{equation}
The Dirac equation (\ref{Dirac0}) then reads \cite{Grib:1994}
\begin{equation} \label{Dirac}
\frac{\ri}{a} \left[ \gamma^0 \frac{\partial}{\partial\eta} + \gamma^1
\frac{\partial}{\partial\chi} + \gamma^2 \frac{1}{f} \frac{\partial}{\partial\theta} +
\gamma^3 \frac{1}{f\sin{\theta}} \frac{\partial}{\partial\phi} + \frac{3a^{\prime}}{2a}
\gamma^0 + \frac{f^{\prime}}{f} \gamma^1 + \frac{\cot \theta}{2f} \gamma^2 \right] \psi -
m\psi = 0 \, ,
\end{equation}
where the prime denotes derivative with respect to the conformal time $\eta$.
%Presenting
%its solution in the form
%\begin{equation}
%\psi (x) = a^{-3/2} \left[ f_{k+}(\eta)I\oplus f_{k-}(\eta)I \right] N_k
%(r, \theta, \phi) \, ,
%\end{equation}
%where $I$ is the unit spin matrix, $k$ is the comoving wave number, and $N_k (r, \theta,
%\phi)$ are canonically normalized time-independent bispinors, one gets the following
%equation for time functions $f_{k \pm} (\eta)$\,:
%\begin{equation}
%f'_{k \pm} + \ri k f_{k \mp} \pm \ri m a f_{k \pm} = 0 \, .
%\end{equation}

Let us consider the case of pseudo-Euclidean space with $f=\chi$. In this case, it is
convenient to use the Cartesian coordinates. Then equation (\ref{Dirac}) becomes
\beq
\label{DiracN}
\left[\frac{\ri}{a}\gamma^{\mu}\partial_{\mu}+\frac{3i}{2a}\frac{a^{\prime}}{a}\gamma^0-m(\eta)\right]\psi=0
\, .
\eeq

In order to separate variables in this equation, one can choose the solution in the form
\begin{equation}
\psi_j^{(\pm)}=\frac{1}{(2\pi a)^{3/2}}u^{(\pm)}_{k\sigma}(\eta)e^{\pm ik_{\alpha}x^{\alpha}}
\end{equation}
Here, the collective quantum number $j=\{k,\sigma\}$, with
$k=\left\{k_1,k_2,k_3\right\}$, $\sigma=\pm1$, and
\begin{equation}
u^{+}_{\sigma k}=\left(\begin{array}{c}\sigma
f_{k+}(\eta)\varphi_{\sigma}(k)\\-f_{k-}(\eta)\varphi_{\sigma}(k)\end{array}\right),\quad
u^{-}_{\sigma k}=\left(\begin{array}{c}f_{k+}(\eta)\varphi_{\sigma}(k)\\\sigma
f_{k-}(\eta)\varphi_{\sigma}(k)\end{array}\right) \, .
\end{equation}
The spinors $\varphi_{\sigma}$ are the eigenfunctions of the helicity operator
\begin{equation}
\left(\vec{\sigma} \vec{n}\right)\varphi_{\sigma}=\sigma\varphi_{\sigma} \, ,
\end{equation}
where $\vec{n} = \vec{k} / | \vec{k} |$, and $\vec{\sigma}$ are the Pauli matrices.
Then, for the first component of equation (\ref{DiracN}), one has
\begin{equation}
\frac{\ri}{a}\left(\vec{\sigma} \vec{n}\right)\frac{\ri k}{(2\pi
a)^{3/2}}f_{k-}\varphi_{\sigma}+\left(\frac{\ri}{a} \frac{\partial}
{\partial\eta}+\frac{3\ri}{2a}\frac{a^{\prime}}{a}-m\right)\left(\frac{\sigma}{(2\pi
a)^{3/2}}f_{k+}\varphi_{\sigma}\right)=0 \, ,
\end{equation}
getting, for the time-dependent functions,
\begin{equation}
f_{k+}^{\prime}+ \ri k f_{k-}+\ri m a f_{k+}=0,
\end{equation}
and
\begin{equation}
f_{k-}^{\prime} + \ri k f_{k+} - \ri m a f_{k-}=0\, .
\end{equation}

As in the scalar case, one can express the general solution of this system in terms of
the Bogolyubov coefficients $\alpha_k$ and $\beta_k$ as follows:
\begin{equation}
f_{k \pm} (\eta) = {} \pm N_{\mp} (\eta) \alpha_k (\eta) e^{\ri \int_{\eta_0}^\eta \Omega_k (\eta') a
(\eta') d \eta'} - N_{\pm} (\eta) \beta_k (\eta) e^{- \ri \int_{\eta_0}^\eta \Omega_k (\eta') a
(\eta') d \eta'} \, ,
\end{equation}
with
\begin{equation}
N_{\pm} = \sqrt{\frac{\Omega_k \pm m}{\Omega_k}}\, , \qquad \Omega_k^2 = \frac{k^2}{a^2} + m^2 \, .
\end{equation}
The Bogolyubov coefficients satisfy the following system of equations:
\ber
\beta'_k &=& \frac{2 m a' + m' a}{2 k} e^{2 \ri \int_{\eta_0}^\eta \Omega_k (\eta') a (\eta') d \eta'}
\alpha_k \, , \\
\alpha'_k &=& \frac{2 m a' - m' a}{2 k} e^{-2 \ri \int_{\eta_0}^\eta \Omega_k (\eta') a
(\eta') d \eta'} \beta_k \, .
\eer

Proceeding to the usual cosmological time $t$, we can just repeat the derivation and the
arguments of the preceding section with $\phi (t)$ in (\ref{varmass}) given by
(\ref{oscill}). Eventually, we obtain the first-order perturbation-theory solution for
the coefficient $\beta_k$ in the form
\begin{equation}
\beta_k = \frac{\Upsilon M}{4 \ri} \int_{t_0}^t \frac {\phi_0 (t') }{\omega_k (t')} e^{2
\ri \int^{t'}_{t_0} \omega_k (t^{\prime\prime}) d t^{\prime\prime} - \ri M t'} d t' \, ,
\end{equation}
where
\beq
\omega_k = \sqrt{m_\psi^2 + \frac{k^2}{a^2}} \approx \frac{k}{a} \, .
\eeq
Higher-order corrections to this perturbative solution are small under the condition
\cite{Shtanov:1994ce}
\beq \label{pertf}
\frac{\Upsilon \phi_0}{M} \ll 1 \, ,
\eeq
which is assumed to be the case.

By using the stationary-phase approximation, we obtain, similarly to (\ref{ocnumb}),
\beq\label{ocnumf}
\beta_k = - \frac{\Upsilon M \phi_0 (t_k)}{4 \omega (t_k)} \sqrt{\frac{\pi}{| \dot
\omega_k (t_k) |}} = - \frac12 \Upsilon \phi_0 (t_k) \sqrt{\frac{\pi}{| \dot \omega_k
(t_k) |}} = - \frac12 \Upsilon \phi_0 (t_k) \sqrt{\frac{2 \pi}{M H (t_k)}} \, .
\eeq
Then, repeating the reasoning of the end of Sec.~\ref{sec:bose}, and taking into account
the four spin polarizations of particles and anti-particles, we get the final result for
the production rate of fermions $\Gamma_\psi$, which coincides with (\ref{Gammaf}).

\section{The Starobinsky model}
\label{sec:starobinsky}

Historically, one of the first models that exhibited inflation was the model suggested by
Starobinsky \cite{Starobinsky:1980te}.  It is motivated by the necessity to consider
local quantum corrections to the Einstein theory of gravity. The simplest such correction
represents the term proportional to the second power of the Ricci scalar in the action of
the model, so that the full gravitational action reads as
\begin{equation}\label{action-g}
S_\rg = -\frac{M_\rP^2}{2} \int d^4 x \sqrt{-g} \left( R - \frac{R^2}{6\mu^2} \right)\, ,
\end{equation}
where
\begin{equation}\label{2a}
\mu = 1.3 \times 10^{-5} M_\rP
\end{equation}
is a constant with indicated value required to explain the inflationary origin of the
primordial perturbations \cite{Faulkner:2006ub}.

The free scalar ($\varphi$) and spinor ($\psi$) fields are described by the usual actions
\ber
S_\varphi &=& \frac12 \int d^4 x \sqrt{-g} \left( g^{\mu\nu} \partial_\mu \varphi
\partial_\nu \varphi - m_\varphi^2 \varphi^2 \right) \, , \label{scal} \\
S_\psi &=& \int d^4 x \sqrt{-g} \left( \ri \bar \psi \cD \psi - m_\psi \bar \psi \psi
\right) \, . \label{spin}
\eer

A conformal transformation $g_{\mu\nu} \to \chi^{-1} g_{\mu\nu}$ with
\beq
\chi = \exp \left(\sqrt{\frac23} \frac{\phi}{M_\rP} \right)
\eeq
transforms the theory (\ref{action-g}) into the usual Einstein gravity with a new special
scalar field $\phi$\,:
\begin{equation}\label{6}
S_\rg = -\frac{M_\rP^2}{2} \int d^4 x \sqrt{- g}
R + \int d^4 x \sqrt{- g} \left[ \frac12 \, g^{\mu\nu} \partial_\mu \phi \partial_\nu
\phi - V(\phi) \right]\, ,
\end{equation}
where
\begin{equation}\label{scalpot}
V(\phi) = \frac{3\mu^2 M_\rP^2}{4} \left[1 - \chi^{-1} (\phi) \right]^2
\end{equation}
is the arising field potential.

The new scalar field $\phi$ (called the {\em scalaron\/}) universally interacts with
other fields present in the theory. Thus, after the appropriate conformal transformation
$\varphi \rightarrow \chi^{1/2}\varphi$, $\psi \rightarrow \chi^{3/4}\psi$, and $\cD
\rightarrow \chi^{1/2}{\cD}$, the actions for scalar and fermion fields take the form,
respectively,
\begin{equation} \label{action_b}
S_{\varphi}=\int d^4x \sqrt{-g} \left(\frac12 g^{\mu\nu} \partial_\mu \varphi \partial_\nu
\varphi - \frac{1}{2\chi} m_{\varphi}^2 \varphi^2 + \frac{\varphi^2}{12M_\rP^2}
g^{\mu\nu} \partial_{\mu} \phi \partial_{\nu} \phi + \frac{\varphi}{\sqrt{6}M_\rP}
g_{\mu\nu} \partial_{\mu} \varphi \partial_{\nu} \phi \right) \, ,
\end{equation}
\begin{equation} \label{action_f}
S_\psi = \int d^4 x \sqrt{- g} \left(\ri \bar \psi \cD \psi -
\frac{m_\psi}{\sqrt{\chi}} \bar \psi \psi \right) \, .
\end{equation}

The inflationary model based on the scalaron $\phi$ is quite successful in solving the
problems of the Big-Bang theory and is consistent with modern observations
\cite{Gorbunov:2010bn}. In this model, after the end of inflation, the scalaron starts
oscillating near the minimum of its potential (\ref{scalpot}), which leads to production
of particles in the external field of the oscillating scalaron. During most part of this
stage, the condition
\beq \label{smallamp}
\left| \frac{\phi}{M_\rP} \right| \ll 1
\eeq
is valid, and, as the scalaron amplitude decreases, this inequality becomes stronger with
time. Without taking into account the back-reaction of the mater fields on the dynamics
of the scalaron field, its behavior at the stage of preheating is then approximately
described by the Klein--Gordon equation
\beq \label{eq-phi}
\Box \phi + \mu^2 \phi = 0 \, ,
\eeq
and by the oscillatory regime (\ref{oscill}) with mass $M = \mu$.

In this section, we are going to show that post-inflationary particle production in the
model under consideration is well described by the theory developed in the preceding
sections.

The equation of motion for the scalar $\varphi$ field follows from (\ref{action_b}):
\beq
\Box \varphi + \left[ \chi^{-1} (\phi)\, m_\varphi^2 + \frac{\Box \phi}{\sqrt{6} M_\rP} -
\frac{\left(\partial \phi \right)^2}{6 M_\rP^2} \right] \varphi = 0 \, ,
\eeq
or, using the equation of motion (\ref{eq-phi}) of the scalaron in the neighborhood of
its minimum,
\beq \label{eq-varphi1}
\Box \varphi + \left[ \chi^{-1} (\phi)\, m_\varphi^2 - \frac{\mu^2 \phi}{\sqrt{6} M_\rP}
- \frac{\left(\partial \phi \right)^2}{6 M_\rP^2} \right] \varphi = 0 \, .
\eeq

Due to condition (\ref{smallamp}), the last term in the brackets of (\ref{eq-varphi1}) is
much smaller than the previous term, so we can drop it. Assuming also that $m_\varphi \ll
\mu$, we obtain an approximate equation
\beq
\Box \varphi + \left[ m_\varphi^2 - \frac{\mu^2 \phi}{\sqrt{6} M_\rP} \right] \varphi = 0
\, .
\eeq
This is just equation (\ref{eq-varphi}) with
\beq \label{sigma-real}
\zeta = - \frac{\mu^2}{2 \sqrt{6} M_\rP} \, ,
\eeq
and one can check that condition (\ref{smallss}) is satisfied.

Thus, the theory of Sec.~\ref{sec:bose} is applicable here, and the particle production
rate is given by (\ref{Gammab}):
\begin{equation}
\Gamma_{\varphi} = \frac{\zeta^2}{8 \pi M} = \frac{\mu^3}{192 \pi M_\rP^2} \, ,
\end{equation}
which coincides with equation (7) of \cite{Gorbunov:2010bn}.

The spinor field $\psi$, under condition (\ref{smallamp}), interacts with the scalaron
field $\phi$ through the Yukawa coupling (\ref{Yukawa}) with strength
\beq \label{yukawa-real}
\Upsilon = \sqrt{\frac23} \frac{m_{\psi}}{M_\rP} \, .
\eeq
Again, one can see that the condition (\ref{pertf}) for perturbative solution is
satisfied due to the conditions $m_\psi \ll M = \mu$ and (\ref{smallamp}). We then
calculate the relevant decay rate (\ref{Gammaf}):
\begin{equation}
\Gamma_{\psi} = \frac{\Upsilon^2 M}{8 \pi} = \frac{\mu m_\psi^2}{12 \pi
M_\rP^2} \, ,
\end{equation}
which, up to a factor four (possible account of the spin states), coincides with equation
(8) of \cite{Gorbunov:2010bn}.

\section{Inflaton production}
\label{sec:inflaton}

During preheating, inflaton field with nonlinear potential can produce its own
excitations.  To consider such self-production, we separate the inflaton field
$\phi(\textbf{x},t)$ into the background part and excitation:
\begin{equation}
\phi(\textbf{x},t)=\phi(t)+\delta\phi(\textbf{x},t) \, .
\end{equation}

Substituting the field in this form into the equation of motion
\begin{equation}
\ddot{\phi}-\nabla^2\phi+ V^{\prime}_{\phi}=0 \, ,
\end{equation}
one gets the equations for the background field and its excitation:
\begin{equation}
\ddot{\phi}(t)+V^{\prime}_{\phi}(t)=0\, ,
\end{equation}
\begin{equation}
\delta\ddot{\phi}_{k} (\textbf{x},t) + k^2 \delta \phi_{k} (\textbf{x},t) +
V^{\prime\prime}_{\phi\phi}\delta\phi_{k}(\textbf{x},t)=0 \, .
\end{equation}
Here, we have used the expansion
\begin{equation}
V^{\prime}(\phi+\delta\phi)=V^{\prime}(\phi)+V^{\prime\prime}(\phi)\delta\phi
\end{equation}
and proceeded to the Fourier space.

At the reheating stage in the Starobinsky model, for $\phi/M_\rP \ll 1$, we have
\begin{equation}
V^{\prime\prime}=\mu^2\left(1-\sqrt{6}\frac{\phi}{M_P}-\frac43\left(\frac{\phi}{M_P}\right)^2\right)\, .
\end{equation}
Then, restricting ourselves to the first order in $\phi / M_\rP$, we obtain
\begin{equation}
\label{fluct}
\delta\ddot{\phi}_{k}+\left(\frac{k^2}{a^2}+\mu^2-\sqrt{6}\mu^2\frac{\phi}{M_P}\right)\delta\phi_{k}=0 \, .
\end{equation}
This equation is similar to (\ref{basic}) with the background part of inflaton field
behaving as in (\ref{oscill}). Thus, we can threat (\ref{fluct}) as described in
Sec.~\ref{sec:bose}. For the Bogolyubov coefficient, we have the equation
\begin{equation}
\dot{\beta}_k=\frac{\dot{\Omega}_k}{2\Omega_k}e^{2i\int\limits_{t_0}^{t}\Omega_k(t^{\prime})dt^{\prime}}
\end{equation}
with
\begin{equation}
\Omega_k^2=\omega_k^2(t)-\sqrt{6}\mu^2\frac{\phi_0(t)}{M_P}\cos{\mu t}\, ,
\end{equation}
\begin{equation}
\omega_k^2=\mu^2+\frac{k^2}{a^2} \, .
\end{equation}

In the approximation $\phi_0 / M_\rP \ll 1$, we obtain
\begin{equation}
\frac{\dot{\Omega}_k}{2\Omega_k}=\frac{\sqrt{3}}{2\sqrt{2}}\frac{\mu^3\phi_0}{\omega_k^2M_p}\sin{\mu t}
\end{equation}
and
\begin{equation}
\label{e10}
\dot{\beta}_k = \frac{\ri\sqrt{3}}{4\sqrt{2}} \frac{\mu^3}{\omega_k^2}
\frac{\phi_0}{M_P}\left(e^{2 \ri \int\limits_{t_0}^{t}\Omega_k(t^{\prime})dt^{\prime}- \ri \mu
t}-e^{2 \ri \int\limits_{t_0}^{t}\Omega_k(t^{\prime})dt^{\prime}+ \ri \mu t}\right)
\end{equation}
The specific feature of this equation is that the system does not pass through the
resonance.  Assuming that the pre-exponent coefficient in (\ref{e10}) changes slowly with
time compared to the exponent, one can estimate the value of $\beta_k$, integrating only exponent, to obtain
\begin{equation}
|\beta_k|^2 \simeq \frac{3}{32}\frac{\phi_0^2}{M_P^2}\frac{\mu^4}{\omega_k^4} \, .
\end{equation}

For the scalar particles, using (\ref{ocnumb}) and (\ref{sigma-real}), we have
\begin{equation}
|\beta_k|^2=\frac{\pi \mu \phi_0^2}{12M_P^2 H} \, .
\end{equation}
Thus, the ratio of the production rates of the matter scalar particles and inflaton is
approximately $\mu / H$, and, since $H \ll \mu$, the production of inflaton is
negligible. However, it might be comparable with the production of fermions of small mass
with [see (\ref{ocnumf}) and (\ref{yukawa-real})]
\begin{equation}
|\beta_k|^2=\frac{\pi}{3}\frac{m_{\psi}^2}{M_P^2}\frac{\phi_0^2}{\mu H} \, .
\end{equation}

\section{Discussion}
\label{sec:conclude}

Particle production in the background of an external classical oscillating field is one
of the key processes describing the stage of preheating after inflation.  Since the
beginning of 1990s \cite{Traschen:1990sw, KLS, Shtanov:1994ce}, it is known that this
process often cannot be described by the usual decay rates of the inflaton calculated in
perturbation theory.  Thus, in Minkowski space, it would be dominated by the parametric
resonance in the lowest resonance band no matter how small is the coupling between the
inflaton and bosonic matter fields.  The energy density of created particles would grow
exponentially with time, in contrast to the usual perturbation-theory expectations.  The
process of creation of fermions would be described non-perturbatively as well \cite{GK}.

The specific features of the expanding universe, surprisingly, restore the validity of
the usual Born formula in the case of sufficiently small coupling.  The reason is that
every particular mode of the field to be excited spends only a finite amount of time in
the resonance zone, resulting in small occupation numbers. Calculating the rates of
particle production in the stationary-phase approximation, we obtain the standard
classical formulas for the energy transfer between the inflaton and scalar and spinor
fields in the case of small couplings.  They are characterized by the standard effective
decay rates (\ref{Gammab}) and (\ref{Gammaf}), respectively, and do not depend on the
details of the universe expansion.  The reason of this peculiar property can be seen from
equations (\ref{ocnumb}) and (\ref{ocnumf}). The mean occupation numbers of particles are
$N_k = |\beta_k|^2 \propto 1 / H (t_k)$, inversely proportional to the Hubble parameter
at the time when the mode with wave number $k$ passes through the resonance zone and gets
excited.  On the other hand, the rate of energy production is determined by the rate of
filling new modes, which is governed by frequency redshift and is directly proportional
to the Hubble parameter. These two effects compensate for each other leading to a
constant effective values of $\Gamma_\varphi$ and $\Gamma_\psi$.

It is also important that couplings (\ref{scoup}) and (\ref{Yukawa}) were linear in the
inflaton field $\phi$, so that we also had $N_k = |\beta_k|^2 \propto \phi_0^2 \propto
\rho_\phi$. These are precisely the types of coupling responsible for the decay of a
$\phi$-particle, so that interpretation of (\ref{Gammab}) and (\ref{Gammaf}) as decay
rates makes sense.

We have shown that the considerations and results of the present paper are fully
applicable to one of the most successful inflationary models --- the Starobinsky model.

\medskip

S.~V. is grateful to the Swiss National Science Foundation (individual grant
No.~IZKOZ2-154984), to  Prof.\@ Ruth Durrer from the Geneva University and also to Marc Vanderhaeghen and Dr. Vladimir Pascalutsa from the Institut f\"{u}r Kernphysik, Johannes
 Gutenberg Universit\"{a}t Mainz, Germany, where part of this work was done, for their support and kind hospitality. This work was supported in part by the SFFR of Ukraine Grant
No.~F53.2/028 and by the Swiss National Science Foundation grant SCOPE IZ7370-152581.


\begin{thebibliography}{9}

\bibitem{Linde:1990}
A.~D.~Linde, {\em Particle Physics and Inflationary Cosmology\/}, Harwood, Chur,
Switzerland (1990).

\bibitem{Traschen:1990sw}
  J.~H.~Traschen and R.~H.~Brandenberger,
  %``Particle Production During Out-of-equilibrium Phase Transitions,''
  Phys.\ Rev.\ D {\bf 42}, 2491 (1990).
  %%CITATION = PHRVA,D42,2491;%%

\bibitem{KLS}
%\bibitem{Kofman:1994rk}
  L.~Kofman, A.~D.~Linde and A.~A.~Starobinsky,
  %``Reheating after inflation,''
  Phys.\ Rev.\ Lett.\  {\bf 73}, 3195 (1994)
  [hep-th/9405187];\
%\bibitem{Kofman:1997yn}
  L.~Kofman, A.~D.~Linde and A.~A.~Starobinsky,
  %``Towards the theory of reheating after inflation,''
  Phys.\ Rev.\ D {\bf 56}, 3258 (1997)
  [hep-ph/9704452].
  %%CITATION = HEP-PH/9704452;%%

\bibitem{Shtanov:1994ce}
  Y.~Shtanov, J.~H.~Traschen and R.~H.~Brandenberger,
  %``Universe reheating after inflation,''
  Phys.\ Rev.\ D {\bf 51}, 5438 (1995)
  [hep-ph/9407247].

\bibitem{GK}
%\bibitem{Greene:1998nh}
  P.~B.~Greene and L.~Kofman,
  %``Preheating of fermions,''
  Phys.\ Lett.\ B {\bf 448}, 6 (1999)
  [hep-ph/9807339];\
  %%CITATION = HEP-PH/9807339;%%
%\bibitem{Greene:2000ew}
  P.~B.~Greene and L.~Kofman,
  %``On the theory of fermionic preheating,''
  Phys.\ Rev.\ D {\bf 62}, 123516 (2000)
  [hep-ph/0003018].
  %%CITATION = HEP-PH/0003018;%%

\bibitem{Starobinsky:1980te}
  A.~A.~Starobinsky,
  %``A New Type of Isotropic Cosmological Models Without Singularity,''
  Phys.\ Lett.\ B {\bf 91}, 99 (1980).
  %%CITATION = PHLTA,B91,99;%%

\bibitem{Grib:1994}
A.~A.~Grib, S.~G.~Mamayev, V.~M.~Mostepanenko,  {\em Vacuum Quantum Effects in Strong
Fields\/}, Friedmann Laboratory Publishing, St.-Petersburg (1994).

\bibitem{Faulkner:2006ub}
  T.~Faulkner, M.~Tegmark, E.~F.~Bunn and Y.~Mao,
  %``Constraining f(R) Gravity as a Scalar Tensor Theory,''
  Phys.\ Rev.\ D {\bf 76}, 063505 (2007)
  [astro-ph/0612569].
  %%CITATION = ASTRO-PH/0612569;%%

\bibitem{Gorbunov:2010bn}
  D.~S.~Gorbunov and A.~G.~Panin,
  %``Scalaron the mighty: producing dark matter and baryon asymmetry at reheating,''
  Phys.\ Lett.\ B {\bf 700}, 157 (2011)
  [arXiv:1009.2448 [hep-ph]].

\end{thebibliography}
\end{document}